\documentclass{aa}  
\usepackage{graphicx}
\usepackage{txfonts}
\begin{document}
   \title{Accreting corona model of the X-ray variability in soft state 
             X-ray binaries and active galactic nuclei}

   \author{A. Janiuk
          \inst{1,2}
          \and
          B. Czerny\inst{1}
          }

   \offprints{A. Janiuk}

   \institute{Copernicus Astronomical Center, Bartycka 18, Warsaw, Poland\\
         \and
        Department of Physics, University of Nevada, Las Vegas, NV89154, USA\\
             \email{agnes@camk.edu.pl, bcz@camk.edu.pl}
             }

   \date{Received 31/07/2006; Accepted 20/02/2007}

% \abstract{}{}{}{}{} 
% 5 {} token are mandatory
 
  \abstract
   {We develop a two-flow model of accretion onto a black hole which incorporates the
   effect of the local magneto-rotational instability. The flow consists of
   an accretion disk and an accreting corona, and the local dynamo affects
   the disk/corona mass exchange.}
   {The model
   is aimed to explain the power spectrum density of the sources in their soft,
   disk-dominated states.}
   {The local perturbations of the magnetic field in the disk are 
   described as in King et al. (2004) and Mayer and Pringle (2005), but
   the time-dependent local magnetic field is assumed to affect the local
   supply of the material to the corona. Since the viscous timescale in the 
   corona is much shorter than in the
   disk, the local perturbations are not smeared in the corona.
   Simple analytical estimates and
   full time-dependent computations of the disk-corona system are performed.}
   {The accreting corona model can reproduce the broad power spectra of Soft State X-ray binaries and AGN. The model, however, predicts that (i) sources undergoing radiation pressure instability (GRS 1915+105) should have systematically steeper power spectra than other sources, (ii) AGN should have systematically steeper power spectra than GBH, even if their disks are described using viscosity proportional to the gas pressure. More precise measurements of
power spectra of Soft State sources are clearly needed.}
   {}

   \keywords{accretion disks --
                X-rays:binaries --
                galaxies:active
               }

  \authorrunning{A. Janiuk and B. Czerny}
  \titlerunning{Accreting corona model of the X-ray variability}
   \maketitle
%
%________________________________________________________________

\section{Introduction}

The X-ray emission of the Galactic Black Holes (GBH) and active galactic
nuclei (AGN) is strongly variable. Aperiodic character of this variability
(GBH: Terrell \cite{terrel72}, for a review see van der Klis \cite{klisrev},
Remillard \& McClintock \cite{remillard06}; AGN: McHardy \& Czerny \cite{fractal}, Lawrence et al. \cite{lawrence}; for
a review, see Mushotzky et al. \cite{mushotzky}) makes the interpretation of the physical nature of the phenomenon rather
difficult. On the other hand, several 
important properties of this variability were found: (i) the shape of the 
power spectrum depends significantly on the luminosity state of the system 
(e.g. Pottschmidt et al. \cite{pott2003}, McHardy et al. \cite{mchardy2004}); 
(ii) in Soft State sources the slope of the power spectrum is close to 1 in
a broad frequency band, only steepening off at the highest frequencies which
means that a broad range of frequencies contribute roughly equal power to the
overall variability (e.g. Revnivtsev et al. \cite{revni2000});
 (iii) in Hard State sources the power spectrum is more
complex (although the energy spectrum is much simpler, being roughly of a power
law shape with energy index $\sim 0.9$), well modeled by a few Lorentzians 
with a power law component also present (e.g. Pottschmidt et al. \cite{pott2003}); 
(iv) the simplest shot noise cannot account for the brightest flares unless the flares come in bunches or avalanches (Canizares \& Oda \cite{canizares77}, Negoro et al. \cite{negoro95}, Poutanen \& Fabian \cite{poutanen99}); 
(v) the un-normalized excess variance 
scales with the flux (see e.g. Uttley \& McHardy \cite{uttley2001}). The Hard State sources are possibly filtered versions of
the Soft State variability, therefore the understanding of the Soft State 
variability is of key importance. 

Several models of X-ray variability were considered: rotating hot spots at the
accretion disk surface (Bath et al. \cite{bath74}, Abramowicz et al. 
\cite{abramowicz92}), magnetic flares 
above the accretion disk (Galeev et al. \cite{galeev79}, Poutanen \& Fabian \cite{poutanen99}, Goosmann et al. \cite{goosmann2006}), variable obscuration (e.g.
Boller et al. \cite{boller97}, McKernan \& Yaqoob \cite{mckernan}, Abrassart \& Czerny \cite{abrassart}, Karas et al. \cite{karas2000}),
 local turbulence (Nowak \& Wagoner \cite{nowak95} ). But the key problem
in these models remained: why long timescale variability, plausibly related
to the processes in the outer parts of the disk contributes so significantly
to the overall variability although the energy dissipated in the outer parts
of the disk is very low? 

A promising solution was suggested by Lyubarskii (\cite{Lyubarskii}) 
who postulated that 
the basic variability mechanism operates locally at each disk radius, as
fluctuations of the viscosity parameter $\alpha$ caused by the dynamo mechanism, 
and these perturbations propagate inwards, finally modulating the accretion
rate in the innermost part of the disk. Therefore, these modulations, 
containing the 
memory of the whole disk perturbations are finally seen as X-ray emission 
variability. Such a mechanism in a natural way explains the slope of the power
spectrum in a disk extending down to the marginally stable orbit.

However, the propagation of the perturbations, as envisioned in Lyubarskii 
model, is not a straightforward issue. In the original model, local dynamo
effect was expected to generate local fluctuations in the accretion rate, and
these fluctuations were expected to propagate inwards. The problem is that
these fluctuations propagate in a viscous time, and they are smeared unless
the local dynamo timescale is longer than the local viscous timescale, as 
pointed out by Churazov et al. (\cite{chura2001}; see also Kotov et al. 
\cite{kotov2001}, Gleissner et al. 
\cite{gleissner}). These
problems were overcome by combining the local magnetic dynamo with magnetic 
field diffusion and wind/jest outflow (King et al. \cite{king}, Mayer et al.
\cite{mayer}).

In the present paper we propose to accommodate the basic picture of Lyubarskii
within the frame of the accreting corona model. In our model the variations
due to the local dynamo modify the accretion rate within the corona.
%and since the viscous timescale in the corona is much shorter than the viscous
%timescales within the disk, the perturbations are as easily smeared as in the 
%case of the disk propagation. 
The disk perturbations 
propagate in its 
viscous timescale and therefore are smeared, but the viscous timescale in the corona 
is much shorter
than in the disk. This allows the coronal perturbations to be observed.
The model accounts for the correct power spectrum
of the X-ray lightcurves of sources in the soft states as well as for the fact 
that in GBH disk flux seems to remain almost constant  
and all the variability is 
likely coming from the corona (Churazov et al. \cite{chura2001}). 

The structure of the paper is the following: the simple semi-analytical model
is discussed in Sect.~\ref{sec:analytic}, the results based on numerical model of the accreting 
disk - accreting corona system are described in Sect.~\ref{sect:numerical},  
the discussion in the context of the
observational data is presented in Sect.~\ref{sect:discussion}, and we conclude in Sect.~\ref{sect:conclusions}

\section{Semi-analytical model of accreting corona}
\label{sec:analytic}
 
We consider here the case of an accretion disk extending down to the marginally
stable orbit, surrounded by the hot optically thin corona. We postulate that 
the corona itself can also transport the angular momentum and can exchange 
mass and angular momentum with the underlying optically thick disk. The model 
was
proposed in the context of disk evaporation in cataclysmic variables (Meyer \&
Meyer-Hoffmeister \cite{syfon}), and subsequently developed in a number of 
papers (\. Zycki et al. \cite{zycki95}, Witt et al. \cite{witt97}, 
Janiuk et al. \cite{janiukadv}, R\' o\. za\' nska \& 
Czerny \cite{rozanska2000},
Meyer-Hofmesister \& Meyer \cite{meyer2001}, Misra \& Taam \cite{misra}, 
Meyer-Hofmeister et al. \cite{ema2005}). 

Here we simplify the model by assuming no condensation of the coronal material onto the disk, i.e. the disk material is continuously ejected into the corona, according to the adopted laws. 

\subsection{Power law trends}

We will start with simple analytical estimates of the power spectrum of the X-ray lightcurve. This is possible if the dependence of the process on the disk radius is in the form of a power law, and the effect of the inner and outer radius can be neglected. There are three key quantities built into the model.

First, we assume that the variability is caused by some process which causes the coherent changes in the accretion disk properties which span certain range of radii, $\Delta r$. The size of the region is likely to depend on the location of the perturbation. We assume
\begin{equation}
\Delta r \propto r^{\beta_r}.
\end{equation} 
Next, the characteristic timescale of the process depends on the location of the perturbation as 
\begin{equation}
t \propto r^{\gamma_r}.
\end{equation}
Finally, the amplitude of the ejection rate from the unit disk surface element, $\dot m_z$, is also likely to depend on the location. We assume
\begin{equation}
\dot m_z \propto r^{\alpha_r}.
\end{equation}

In such a model, the ejection of mass in a single event has an amplitude $A \propto 
\dot m_z r \Delta r$, if the event happens in a whole ring, and the slope of the power spectrum describing the variability is then given by
\begin{equation}
P(f) \propto f^{-p};~~~~ p = 1 + {3 + 2 \alpha_r + \beta_r \over \gamma_r}.
\label{eq:slopes}
\end{equation}

If we want to reproduce the slope $ p = 1$ characteristic of the Soft State X-ray sources, we should choose models satisfying the condition
\begin{equation}
2 \alpha_r + \beta_r = - 3,
\end{equation} 
and the condition is independent from $\gamma_r$, i.e. the way how the timescales scale down with the radius is
 unimportant, as stressed by Lyubarskii (1997). 

This analysis neglects the propagation time in the corona, and the effect of the boundary conditions.
 
\subsection{Physical semi-analytical model}

We now develop a more complex, semi-analytical approach which nevertheless captures the essential physics of the accreting corona model.
We assume that the dynamo mechanism is operating within the disk and we model
the stochastic aspect of this process as suggested by 
King et al. (\cite{king}).

We construct the discrete radial grid $r_k$ corresponding to the radial size
of the perturbation. We consider two cases: either $\Delta r = H$ (case a), 
where $h$
is the local disk thickness taken from a stationary model, or $\Delta r = r$
(case b),
which represents the scaling with the thickness of the accreting corona. Here
$\Delta r$ corresponds to the difference between the consecutive radii, 
$r_{k+1} - r_k$. 

At each disk radius $r_k$ we create a time series assuming an underlying 
Markoff process
\begin{equation}
u_{n+1} = -\alpha_1 u_n + \epsilon_n,
\label{eq:markoff}
\end{equation}
where $\epsilon_n$ is a random variable between $-\sqrt{3}$ and $\sqrt{3}$,
which has a uniform distribution, zero mean and the dispersion 
equal to unity.
We assume $\alpha_1 = 0.5$ after King et al. (\cite{king}) and 
Mayer \& Pringle
(\cite{mayer}).

The time-step unit at each radius is proportional to the dynamical (orbital) time at a given disk radius:
\begin{equation}
\tau_{mag} = k_d \sqrt{{r^3 \over GM}},
\end{equation}
where $M$ is the black hole mass, and the factor $k_d$ is the parameter of the
model. The parameter $k_d$ can be considerably larger than 1 since the 
magnetic processes are slowed down by macroscopic resistivity (see e.g. 
Tout \& Pringle \cite{tout92}, Stone et al. \cite{stone96}, 
Komissarov et al. \cite{komissarov}).

The Markoff chain formed at each radius describes the time evolution of 
the magnetic field
\begin{equation}
B = u_n B_{max},
\end{equation}
where $B_{max}(r)$ is the variability amplitude of the magnetic field. 
This local normalization factor $B_{max}(r)$ is determined by the effective
viscosity parameter $\alpha$ of Shakura \& Sunyaev (\cite{shakura})
\begin{equation}
{B^2_{max} \over 4 \pi} = \alpha P,
\label{eq:magnet}
\end{equation} 
where $P$ is the total (gas plus radiation) pressure within the disk 
interior.

Now we estimate the amount of outflow of the disk material into the corona, $\dot m_z$. 
We postulate that the local disk magnetic field is responsible for the
outflow of the material from the disk surface to the corona. This outflow
is a combined result of the magnetic field buoyancy and the numerous
possible mechanisms of the coronal heating (e.g. magnetic reconnection, 
Galeev et al. \cite{galeev79}; dissipation of Alfven waves in the corona,
Kuperus \& Ionson \cite{kuper85}; acoustic heating, coupled with convection,
Bisnovatyi-Kogan \& Blinnikov \cite{blin77}; vertical Poynting flux, Merloni
\cite{merloni2003}; electron conduction, Meyer \& Meyer-Hofmeister 
\cite{syfon}, R\' o\. za\' nska \& Czerny \cite{rozanska2000}; ion irradiation, 
Deufel \& Spruit \cite{spruit}) leading to the
disk evaporation. Since the
exact mechanism is under discussion, we adopt three viable parameterizations
of the evaporation process:
\begin{eqnarray}
\nonumber &   {B^2 \over 8 \pi} v_A {\mu m_p \over k T_{vir}} & ~~~{\rm (case~A)} \\
\nonumber \dot m_z = &  {B^2 \over 8 \pi} c {\mu m_p \over k T_{vir}} & ~~~{\rm  (case~B)} \\
\nonumber &   {B^2 \over 8 \pi} v_A {\mu m_p \over k T_{vir}} {1 \over f(r)}  & ~~~{\rm (case~C)} \\
\label{eq:mz}
\end{eqnarray} 
Here we assume that the coronal (ion) temperature is equal to the virial temperature, 
$v_A$ is the Alfven speed ($v_A^2 = P/\rho$, 
where $\rho$ is the density of the disk), and $f(r)$ is the boundary condition factor 
(equal to $1 - \sqrt{3 R_{Schw}/r}$ in the Newtonian case). The dependence of the pressure $P$, 
and the density $\rho$, is taken from the analytical formulae of Shakura \& Sunyaev (1973), 
either for the gas pressure dominated region 
(for opacity dominated by electron scattering), or for the radiation pressure dominated region. 

We further assume that the coronal inflow proceeds in the viscous timescale 
of the corona, which is taken as
\begin{equation}
t_{cor} = {1 \over \alpha_{cor}} \sqrt{{r^3 \over GM}},
\end{equation}
where $\alpha_{cor}$ is the viscosity parameter characterizing the corona.
We allow for the dispersion in a flow of each clump of the coronal material
by a fraction of the inflow timescale, $\delta$. We assume that the
coronal material does not settle down onto a disk, so the effective accretion 
rate at the inner disk radius consists of the coronal clumps ejected from the
disk at each radius under consideration.

Finally, the X-ray luminosity of the flow is assumed to be proportional to
the coronal accretion flow at the inner disk. The lightcurve is analyzed 
using the standard FFT package and the resulting power spectrum is binned for
clarity.

\subsection{Results}
\label{sect:res1}

Since we consider three options for the disk evaporation efficiency, 
two options for the size of the elementary cell, and two cases of the disk structure, 
we finally have a family of 12 models. We summarize them in Table~\ref{tab:analytic}. 

\begin{table}
\caption{Semi-analytical models}
\begin{center}
\begin{tabular}{c c c c  l }     % 5 columns 
\hline\hline       
 Model & $\dot m_z$ & cell & SS model & slope\\ 
\hline
Aa-rad & A & a & radiation pressure & 0.66 (curv) \\
Ba-rad & B & a & radiation pressure & 2.42 \\
Ca-rad & C & a & radiation pressure & 0.48 \\
Ab-rad & A & b & radiation pressure & 1.16 (curv) \\
Bb-rad & B & b & radiation pressure & 2.99 \\
Cb-rad & C & b & radiation pressure & 1.00 \\
Aa-gas & A & a & gas pressure & 1.19 (curv) \\
Ba-gas & B & a & gas pressure & 1.75 \\
Ca-gas & C & a & gas pressure & 1.03 \\
Ab-gas & A & b & gas pressure & 1.15 (curv) \\
Bb-gas & B & b & gas pressure & 1.71\\
Cb-gas & C & b & gas pressure & 1.00\\
\hline                  
\end{tabular}
\end{center}
Evaporation parameterized as in Eq. \ref{eq:mz} (cases A, B, and C);
cells {\it a} and {\it b} refer to the size of the magnetic field perturbation, 
$\Delta r$, equal either to
the disk thickness $h(r)$ or to the current radius, $r$, respectively. 
The disk structure is described as in Shakura \& Sunyaev (1973), 
either in the radiation pressure dominated region or in the gas pressure dominated region, 
but with the electron scattering opacity. 
The slope of the PSD is measured in the low frequency part. The note (curv) means that a certain level of curvature is seen in the long wavelength part of the PSD.
\label{tab:analytic}
\end{table}

We can understand the results by referring to the power law dependencies given by Eq.~\ref{eq:slopes}. For case (A) and case (C), the important factor is the energy flux provided with the Alfven speed to the corona, $B^2/(8 \pi) v_A$. Since we assume the scaling of the magnetic field with the pressure given by Eq.~\ref{eq:magnet}, it can be easily shown with the use of the hydrostatic equation for the disk structure that for a stationary Shakura-Sunyaev disk the amplitude of this factor reduces to 
\begin{equation}
{B^2 \over 8 \pi} v_A \propto F(r) \propto {f(r) \over r^3},
\end{equation} 
where $F(r)$ is the radiation flux of a stationary disk, and $f(r)$ is the boundary condition factor. Therefore, this value does not depend on the disk model, i.e. whether the disk is dominated by the radiation pressure or the gas pressure. The radial shape of the $\dot m_z$ dependence is then given by $\dot m_z \propto r^{-2} f(r)$ for the case (A) and it is just $\dot m_z \propto r^{-2}$ for the case (C). Therefore, case (C) is characterized roughly by a power law dependence with $\alpha_r = -2$, and case (A) shows a departure from a power law behaviour due to the $f(r)$ factor. This means that the inner boundary condition directly affects the shape of the PSD.

\begin{figure}
   \centering
    \includegraphics[width=8.5cm]{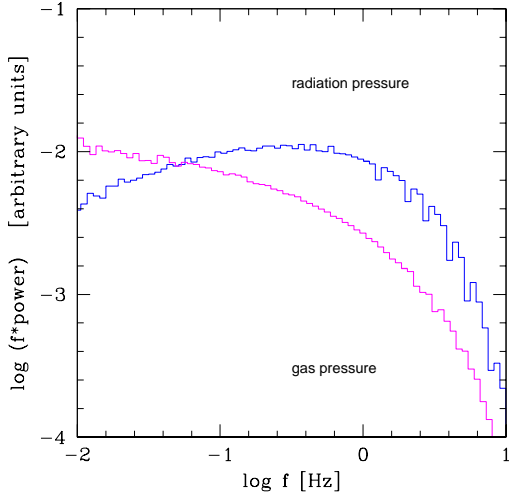}
      \caption{Power spectrum density for the radiation-pressure dominated
    and gas-pressure dominated  
    disk if the dynamo cells scale with the disk
    thickness (models Aa-rad and Aa-gas from Table~\ref{tab:analytic}).  
              }
         \label{fig:pdsradiation}
   \end{figure}

The resulting power spectrum strongly depends on the disk thickness, if
the dynamo cells scale down with it. As an illustration, we show in 
Fig.~\ref{fig:pdsradiation} the two
power spectra: one for radiation pressure dominated disk and one for the gas
pressure dominated disk. This means that the power spectra of AGN with disks
likely to be dominated by radiation pressure should show systematically
smaller values of the slope of the power spectrum, $p$. 

   \begin{figure}
   \centering
   \includegraphics[width=8.5cm]{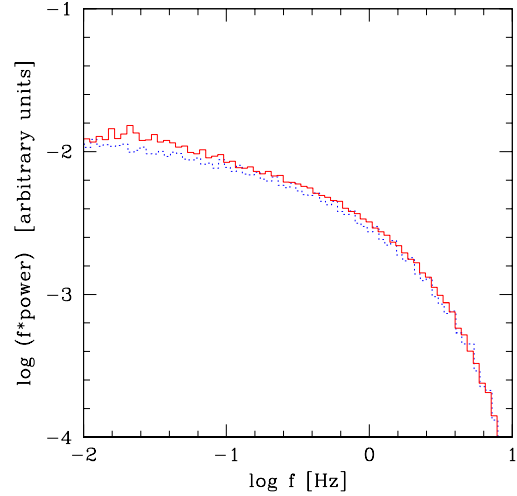}
      \caption{Power spectrum density for the radiation-pressure dominated
     (dotted histogram)
    and gas-pressure dominated (solid histogram) 
    disk if the dynamo cells scale with the disk
    radius. Random location of cells (models Ab-rad and Ab-gas from 
    Table~\ref{tab:analytic}). 
              }
         \label{fig:pds_thick2}
   \end{figure}

However, if we assume that the magnetic cell is large and equal to the
current disk radius ($\Delta r = r$), the power spectrum is the same for the
gas pressure and for the radiation pressure disk model 
(see Fig.~\ref{fig:pds_thick2}). 
There are only a few cells, and if the 
cells were always located at
the fixed disk radii, the power spectrum would show clear regular peaks. 
We may overcome this numerical problem by means of 
randomization of the flare location around the predicted location or
by the assumption of the thinner cells  (e.g. $\delta r = 0.05 r$).
This smears
this effect out and the resulting power spectrum, shown in the Figure, does not show specific
features.
The resulting power spectrum is slightly steeper than $p = 1$. 

Modification of the energy and mass transfer between the disk and the corona 
also affects the PSD slope significantly. Models based on Poynting flux (B-type)
are always very steep, dominated by the longest timescales. C-models, on the
other hand, show no sign of curvature at the long wavelengths, and they are
always flatter. Three of these models have the PSD slope $p$ very close to 1.

\section{Time evolution of the disk with accreting corona}
\label{sect:numerical}

\subsection{Model}

The model of an accretion disk exchanging the mass with the hot
corona is developed from the model described and studied in detail in 
Janiuk \& Czerny (\cite{janiuk2005}).
The basic equations of the (vertically averaged) accretion disk model are:
(i) the viscous heating
\begin{equation}
Q^{+}_{\rm visc} = {3 \over 2}\alpha\Omega H P
\label{eq:heat}
\end{equation}

(ii) the radiative cooling:
\begin{equation}
Q^{-}_{\rm rad}={ P_{\rm rad} c \over \tau}={ \sigma T^{4} \over 
\kappa \Sigma}
\label{eq:qrad}
\end{equation}
where we adopt the electron scattering opacity
$\kappa=0.34$ cm$^{2}$/g;

(iii) the advective cooling (e.g. Paczy\' nski \&
Bisnovatyi-Kogan \cite{bisno}; Muchotrzeb \& Paczy\' nski \cite{mucho}; 
Abramowicz et al. \cite{abra88})
\begin{equation}
Q^{-}_{\rm adv}=
 {2 r P q_{\rm adv} \over 3 \rho GM} F_{\rm tot}
\label{eq:fadv}
\end{equation}
and $q_{\rm adv} \approx 1.0$.

Similarly to the semi-analytical calculations presented in
Section \ref{sec:analytic},  we differentiate between two cases:
gas pressure and radiation pressure dominated disks.
Therefore, we adopt two kinds of heating prescriptions in Eq. 
\ref{eq:heat}: $P = P_{\rm gas}$ and $P = P_{\rm tot}$.
In the latter case the total pressure consists of the gas and radiation pressure,
$P = {k \over m_{\rm p}}\rho T + {1 \over 3} a T^{4}$, and
we use a modified viscosity parameterization
\begin{equation}
 \alpha =\alpha_{0}  {(1+\xi/\xi_{0}) \over (1+(\xi/\xi_{0})^{2} }
\label{eq:visclaw}
\end{equation}
where $\xi = P_{\rm rad}/P_{\rm gas}$ and
$\alpha_{0} = 0.01$ and $\xi_{0}=8.0$ (Nayakshin et al. \cite{melia}).
This is because the constant viscosity parameterization, appropriate for 
the gas pressure dominated disk, would lead to much too large amplitudes
of the disk outbursts due to the thermal-viscous instability induced
by the radiation pressure.

The angular velocity of the disc is assumed to be Keplerian,
$\Omega = \sqrt{GM/r^{3}}$, and the sound speed is $c_{\rm
s}=\sqrt{P/\rho} = \Omega H$. A non-rotating, Schwarzschild black
hole is assumed and the inner radius of the disc is always at 3
$R_{\rm Schw}$. The outer radius is equal to 300 $R_{\rm Schw}$, 
and at this radius
we give a constant mass inflow, parameterized by the external 
accretion rate $\dot M_{\rm ext}$.
The mass of the black hole is assumed to be $10 M_{\odot}$ or 
$ 10^8 M_{\odot}$.

To start the disk evolution, we calculate the initial 
quasi-steady configuration from the
energy balance: $F_{\rm tot} = Q^{+}_{\rm visc} = Q^{-}_{\rm
adv}+Q^{-}_{\rm rad}$, where
the total energy flux $F_{\rm tot}$ dissipated within the disc at a 
radius $r$ is calculated as:
\begin{equation}
F_{\rm tot} = {3 G M \dot M \over 8 \pi r^3} f(r)
\label{eq:ftot}
\end{equation}
and
\begin{equation}
%f(r) = \Big(1-\Big({3 R_{\rm Schw}\over r}\Big)^{3/2}{r-R_{\rm Schw}\over 2 R_{\rm Schw}}\Big)
f(r) = 1 - \sqrt{3 R_{\rm Schw} \over r}
\label{eq:potential}
\end{equation}
is the boundary condition in the Newtonian potential.

The corona  above the disk is thick, $H_{\rm cor}=r$, 
and the pressure in the corona
is its gas pressure due to ions:
$P_{\rm cor} = {k \over m_{\rm p}} \rho_{\rm cor} T_{\rm cor}$.
The coronal temperature is equal to the virial temperature,
$T_{\rm cor} = {G M \over r}{m_{\rm p} \over k}$.

The initial configuration of the corona is computed 
under the assumption that 
the corona has a uniform surface density and 
its optical depth is constant: 
\begin{equation}
\tau_{\rm cor}  = \kappa \Sigma_{cor} = 0.1
\label{eq:taucor}
\end{equation}

The treatment of the mass exchange  in the current model is different 
from the one adopted in the previous paper (Janiuk \& Czerny 2005),
since now we incorporate the stochastic character of the dynamo responsible
for the coronal heating. The mass exchange is proportional
to the locally generated magnetic flux, divided by the energy change per 
particle.
Therefore the rate of the local mass exchange in vertical direction,
in units of g s$^{-1}$ cm$^{-2}$, is given  by Eq.~\ref{eq:mz},
where f(r) is the boundary condition given  by
Eq. \ref{eq:potential}, $v_{\rm A} = \sqrt{\alpha}\Omega H$ is the Alfven velocity and 
the poloidal magnetic field component is calculated from
\begin{equation}
B_{\rm z} = B_{\rm z}^{\rm max} u_{\rm n} = \beta_{\rm S} \sqrt{4\pi\alpha P} u_{\rm n}
\end{equation}
For generality, we introduced here the parameter $\beta_{\rm S} \le 1.0$. 
It is adopted here to account for the possible difference 
between the average magnetic field in the disk and the larger scale magnetic
field likely to be more important for the disk evaporation (e.g. Lovelace, 
Romanova \& Newman \cite{Lovelace}; King et al. \cite{king}). In the previous
section we assumed always $\beta_{\rm S}=1$, for simplicity.

The random changes of the magnetic flux are computed as follows.
First, we assume certain law for the radial distribution of the magnetic cells 
(we consider $\Delta r = H(r)$ - case a, $\Delta r = \sqrt r$ - case b, and $\Delta r = r$ - case c). 
Therefore we impose a separate radial grid to follow
the stochastic changes of the $B$ field in these boxes. Next, we assume that
the timescale of these stochastic fluctuations is proportional to the
dynamical timescale:
\begin{equation}
\tau_{\rm d} = \kappa_{\rm d}/\Omega, 
\label{eq:taudynmag}
\end{equation}
with
$\kappa_{\rm d}=100$.
Since our disc is geometrically thin, with a ratio $H/r$  as small as $\approx 0.01$ 
in some of the small mass models, and even less in the supermassive black hole models, 
the number of points in the magnetic grid varied from $\sim 150$ to $\sim 2250$. 

Having computed the initial disc and corona state we 
solve the equation of mass and
angular momentum conservation:
\begin{equation}
{\partial \Sigma \over \partial t}={1 \over r}{\partial \over \partial
r}(3 r^{1/2} {\partial \over \partial r}(r^{1/2} \nu \Sigma))
- \dot m_{\rm z}
\label{eq:sigevol}
\end{equation}
and the energy equation:
\begin{eqnarray}
{\partial T \over \partial t} + v_{\rm r}{\partial T \over \partial r}
= {T \over \Sigma}{4-3\beta \over 12-10.5\beta}
({\partial \Sigma \over
\partial t}+  v_{\rm r}{\partial \Sigma \over \partial r}) \\
\nonumber +{T\over P H}{1\over 12-10.5\beta} 
(Q^{+}-Q^{-})
\label{eq:tevol}
\end{eqnarray}
Here $\Sigma=H \rho$ is the surface density in the disc, 
$v_{\rm r} = {3 \over \Sigma r^{1/2}} {\partial \over \partial r}
(\nu \Sigma r^{1/2})$
is its radial velocity, and
$\nu=(2P\alpha)/(3\rho\Omega)$ is the kinematic viscosity. 
The heating term is given by Equation \ref{eq:heat}
and the  cooling term $Q^{-}$ is given by Equation \ref{eq:qrad}, 
while the advection is included in the energy equation
via the radial derivatives.

The evolution of the coronal density is also given 
by mass and angular momentum conservation:
\begin{equation}
{\partial \Sigma_{\rm cor} \over \partial t}={1 \over r}{\partial \over \partial
r}(3 r^{1/2} {\partial \over \partial r}(r^{1/2} \nu_{\rm cor} \Sigma_{\rm cor}))
+ \dot m_{\rm z}
\label{eq:sigcorevol}
\end{equation}
The radial velocity in the corona is $v_{\rm r}^{\rm cor} = 
{3 \over \Sigma_{\rm cor} r^{1/2}} {\partial \over \partial r}
(\nu_{\rm cor} \Sigma_{\rm cor} r^{1/2})$ and viscosity is
$\nu_{\rm cor}=(2P_{\rm cor}\alpha_{\rm cor})/(3\rho_{\rm cor}\Omega)$ 
where $\alpha_{\rm cor}=0.01$.
The coronal 
temperature is always equal to the virial 
temperature and does not vary with time.
 
We solve the above set of three time-dependent equations using 
the variables $y=2r^{1/2}$ and $\Xi = y \Sigma$ at the fixed
radial grid, equally spaced in $y$, and
the number of radial zones is set
to $N_{\rm r}=100$. For the time evolution we use the fourth order Runge-Kutta method and the
Adams-Moulton predictor-corrector method. 
We choose the no-torque  inner boundary condition, $\Sigma_{\rm in} =
T_{\rm in} = 0$ for the disc. The outer boundary of the disc is 
parameterized by an
external accretion rate $\dot M_{\rm ext}$. 
The boundary condition in the corona is 
given by Equation \ref{eq:taucor}, with $\tau_{\rm cor}(R_{\rm out}) = 
\tau_{\rm cor}(R_{\rm in}) = 0.1$.

The local mass exchange rate depends on time via the stochastic changes of
the magnetic flux. These are calculated on the magnetic cell's grid 
with $N_{\rm m} \gg N_{\rm r}$. Whenever the time-step exceeds the
local magnetic timescale (Eq. \ref{eq:taudynmag}), the new value of a random number
is created, using the Markoff series according to Eq. \ref{eq:markoff}. 
Because of a different number of points in the two grids, $N_{\rm m}$ and $N_{\rm r}$,
the local mass exchange rate on the basic grid $N_{\rm r}$ can change 0, 1 or a few times
in one time-step. Therefore we simplify this problem, and if the magnetic cells 
are distributed much denser than the basic radial grid points, and $\dot m_{\rm z}$
would change a few times, we take the average of the random numbers $u_{\rm n}$
generated for one point in the same time-step. In particular, 
this is the case in the innermost parts of the accretion disk.

\subsection{Results}

\begin{table}
\caption{Numerical models}
\begin{center}
\begin{tabular}{c c c c c c l }     % 7 columns 
\hline\hline       
 Model & Mass [$M_{\odot}$] & Instab. & $\dot m_z$ & cell size & Heating & slope\\ 
\hline
1 & 10 		& + & A & c & $P_{\rm tot}$ & 1.25 \\
2 & 10 		& + & B & c & $P_{\rm tot}$ & 1.90 \\
3 & 10 		& + & C & c & $P_{\rm tot}$ & 0.99 \\
4 & 10 		& + & A & a & $P_{\rm tot}$ & 1.65 \\
5 & 10 		& + & B & a & $P_{\rm tot}$ & 2.75 \\
6 & 10 		& + & C & a & $P_{\rm tot}$ & 0.89 \\
7 & 10 		& - & A & c & $P_{\rm tot}$ & 1.13 \\
8 & 10 		& - & B & c & $P_{\rm tot}$ & 1.77 \\
9 & 10 		& - & C & c & $P_{\rm tot}$ & 0.85 \\
10 & 10 	& - & A & a & $P_{\rm tot}$ & 1.60 \\
11 & 10 	& - & B & a & $P_{\rm tot}$ & 2.25 \\
12 & 10 	& - & C & a & $P_{\rm tot}$ & 1.37 \\
13 & 10 	& - & A & c & $P_{\rm gas}$ & 1.02 \\
14 & 10 	& - & B & c & $P_{\rm gas}$ & 1.71 \\
15 & 10 	& - & C & c & $P_{\rm gas}$ & 0.38 \\
16 & 10 	& - & A & a & $P_{\rm gas}$ & 0.94 \\
17 & 10 	& - & B & a & $P_{\rm gas}$ & 1.88 \\
18 & 10 	& - & C & a & $P_{\rm gas}$ & 0.58 \\
19 &10$^{8}$& - & A & c & $P_{\rm gas}$ & 1.02 \\
20 &10$^{8}$& - & B & c & $P_{\rm gas}$ & 2.96 \\
21 &10$^{8}$& - & C & c & $P_{\rm gas}$ & 0.82 \\
22 &10$^{8}$& - & A & a & $P_{\rm gas}$ & 0.95 \\
23 &10$^{8}$& - & B & a & $P_{\rm gas}$ & 1.92 \\
24 &10$^{8}$& - & C & a & $P_{\rm gas}$ & 0.78 \\
25 &10$^{8}$& - & C & b & $P_{\rm gas}$ & 0.93 \\
26 &10      & - & C & b & $P_{\rm gas}$ & 0.76 \\
27 &10$^{8}$& - & A & b & $P_{\rm gas}$ & 1.59 \\
28 &10      & - & A & b & $P_{\rm gas}$ & 0.86 \\
29 &10$^{8}$& - & B & b & $P_{\rm gas}$ & 2.97 \\
30 &10      & - & B & b & $P_{\rm gas}$ & 1.72 \\

\hline                  
\end{tabular}
\end{center}
The prescriptions for disk evaporation: A, B, C, as given by Eq. \ref{eq:mz}.
The cell size (a) is $\Delta r \propto H$, 
the size (b) is $\Delta r$, and 
the cell size (c)  $\propto \sqrt{r}$,. 
The viscous heating is proportional either to
the total pressure (with modified viscosity law)
or to the gas pressure. 
Whenever the disk is unstable due to radiation pressure,
and the cell size is proportional to the disk thickness, the
$H(r)$ profile is taken from the hot state of the disk.  
The slope of the PSD is measured in the low frequency part.
\label{tab:numeric}
\end{table}

In Table \ref{tab:numeric} we list the calculated numerical models
and their resulting power spectral slopes.

The slopes were calculated by means of fitting a power law function
to the numerically computed PSD, in the frequency range
$\log f = (-1.8;-0.8)$ Hz and (-8.0; -7.0) Hz
for Galactic black holes and AGN, respectively.
The calculations were performed for stellar mass accreting black hole with $M=10 M_{\odot}$
(gas and radiation pressure dominated disks)
and supermassive black hole with $M=10^{8} M_{\odot}$ (only gas pressure dominated disk).
We adopt the parameterizations of the evaporation process as listed in
Eq. \ref{eq:mz} (cases A, B and C), with different sizes of
the magnetic cell: (a) $\Delta r \propto H$, or (b) $\Delta r \propto r$.
We also calculate the models with size of magnetic cell proportional 
to the square root of 
radius: (c) $\Delta r \propto \sqrt{r}$, which results from our radial grid. 

\begin{figure}
   \centering
   \includegraphics[width=8.5cm]{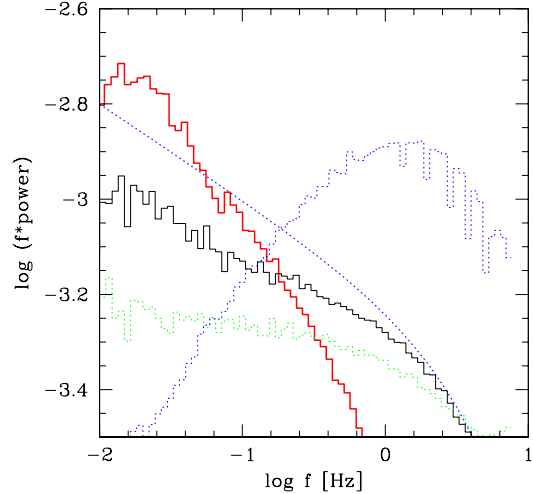}
      \caption{Thick histogram: PSD from the full evolution
      of a disc-corona system with the mass exchange (model 12 from 
     Table~\ref{tab:numeric}) ; dotted histograms: power spectra from 
     semi-analytical
     scheme (lower histogram: model Ca-gas, upper histogram: model Ca-rad); thin lower histogram: power 
     spectrum from semi-analytical model but with numerical prescription 
     for the disk 
     thickness and the total pressure distribution from a stationary numerical model; dotted line is  
representation by a power of the slope 1.2 and the high energy cut-off 10 Hz. 
              }
         \label{fig:pdscompar}
   \end{figure}

The power spectrum resulting from the full time evolution study of the 
disk/corona system systematically differs from the simplified semi-analytical 
model. We can see from Fig.~\ref{fig:pdscompar} that the analytical model with
the same physical assumptions for the disk evaporation rate gives much flatter power spectrum 
(dotted histograms) than the numerical model. 

\begin{figure}
   \centering
   \includegraphics[width=8.5cm]{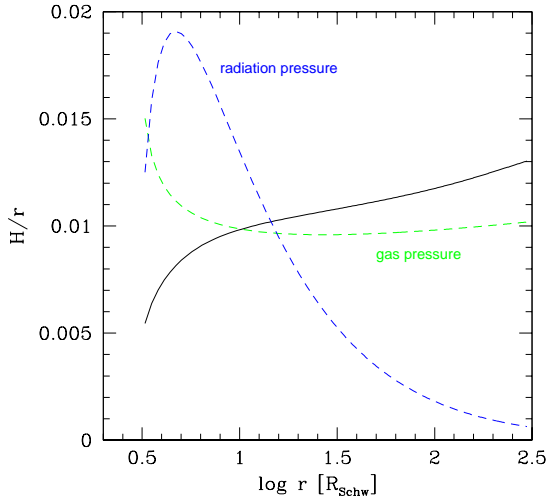}
      \caption{The ratio of the disk thickness to the radius for numerical
    model (continuous line) and analytical radiation pressure and gas pressure 
  models (dashed lines)  shown 
    in Fig.~\ref{fig:pdscompar}.
              }
         \label{fig:ha}
   \end{figure}

One of the reasons is that the disk properties in the numerical model differ
from the analytical prescription. Numerically calculated disk under assumption of the total 
pressure is not fully radiation-pressure dominated, therefore neither the total pressure, 
nor the disk thickness in analytical model actually follow the numerical results. As an example,
we plot in Fig.~\ref{fig:ha} the ratio of the disk thickness in the numerical 
model and in the analytical gas-dominated as well as radiation-pressure dominated disk model. 
Since in the size of the cells in (a) type models for a galactic object 
is rising faster than linearly with
the radius, the variability at the outer edge gains systematically additional 
power with respect to the variations in the inner part of the disk and the 
power spectrum resulting from numerical model is steeper than from the 
analytical one. 
Therefore we also calculated a semi-analytical model but
with the disk thickness and the total pressure distribution taken from a 
numerical stationary model. The result is shown in Fig.~\ref{fig:pdscompar} 
as a thin histogram. The PSD slope of this model is $p=1.13$, i.e. in the 
numerical model the PSD is slightly steeper due to the exact computation of 
the propagation of perturbations through the corona. 

The numerical computations include one more extremely important effect: they
take into account the non-stationarity of some disk models. It is well known
that the heating proportional to the total pressure 
leads to the disk instabilities and luminosity outbursts, if only the
external accretion rate is large enough to make the radiation pressure 
dominate (Pringle, Rees \& Pacholczyk \cite{radinstab}, Shakura \& 
Sunyaev \cite{ssinstab}; see also Szuszkiewicz \& Miller 
\cite{szusz97,szusz98}, Nayakshin et. al. \cite{melia}, Janiuk, Czerny \& 
Siemiginowska \cite{janiuk2000,janiuk2002}, Mayer \& Pringle \cite{mayer}). 
In our models the instability switches on for $\dot m \ge 0.03$, 
for $M = 10 M_{\odot}$ black hole mass. 
In this case the disk
thickness strongly fluctuates in the unstable region. It means that the 
cell size also strongly fluctuates, however numerically it is too much time-consuming 
to follow this evolution at each time step.
Therefore, when we adopt the cell size as proportional to the disk thickness, the
$H(r)$ profile is taken from the hot state solution.

\begin{figure}
   \centering
    \includegraphics[width=8.5cm]{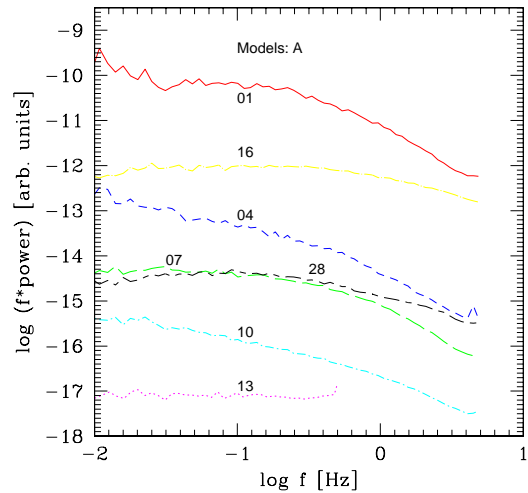}   
      \caption{Power spectra calculated from numerical models
with the prescription $A$ for mass evaporation rate (cf. Eq. \ref{eq:mz}
and Tab. \ref{tab:numeric}; the labels on separate curves refer to the
number of the model as given in the Table).
              }
         \label{fig:pds_A}
   \end{figure}

As in the semi-analytical models, the predicted variability significantly depends on the adopted evaporation law.
In Figure \ref{fig:pds_A}
we plot the power spectra calculated from numerical models
with the prescription $A$ (see Eq.~\ref{eq:mz}).
The total pressure models  "A" with the cell size proportional to $\sqrt{r}$ (01, 07) 
have the power spectra substantially flatter than the models with cell's
of sizes proportional to the disk height (04, 10), respectively
for other parameters kept the same, and the unstable models (01, 04) are steeper than their stable equivalents (07,10). Gas pressure models (13 and 16) have flatter power spectra than the total pressure models.

\begin{figure}
   \centering
  \includegraphics[width=8.5cm]{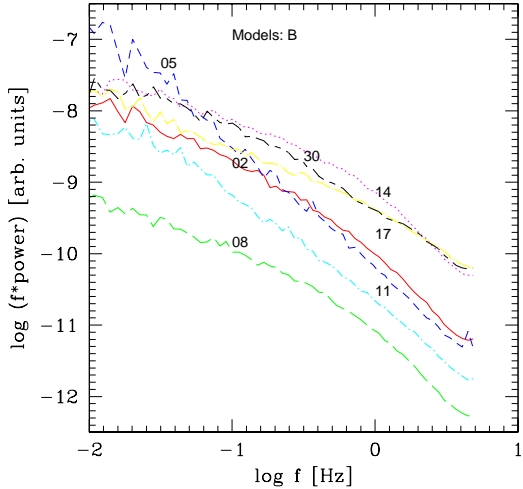} 
      \caption{The same as in Fig \ref{fig:pds_A}, but 
for  the prescription $B$ for mass evaporation rate.
              }
         \label{fig:pds_B}
   \end{figure}

In Figure \ref{fig:pds_B}
we plot the power spectra calculated from numerical models
with the prescription $B$ for mass evaporation rate.  All power spectra are very steep, as in analytical solutions, independently from other assumptions.

\begin{figure}
   \centering
    \includegraphics[width=8.5cm]{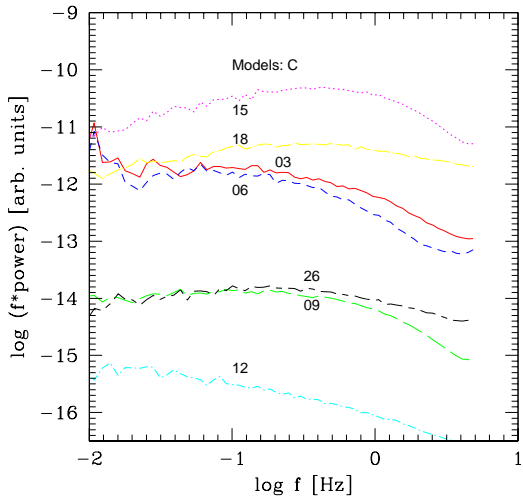}
      \caption{The same as in Fig \ref{fig:pds_A}, but 
for  the prescription $C$ for mass evaporation rate.
              }
         \label{fig:pds_C}
   \end{figure}

In Figure \ref{fig:pds_C}
we plot the power spectra calculated from numerical models
with the prescription $C$ for mass evaporation rate.
This prescription gives the flattest power spectra. The trend of
spectrum flattening for the cell's size proportional 
to $\sqrt{r}$ is present in stable models.
The dependence on the heating term ($P_{\rm tot}$ vs. $P_{\rm gas}$)
is similar to the case of models ``$A$''.

We also calculated the disk-corona system evolution and power spectra for the
case of accreting supermassive black hole in AGN. We took the black hole mass 
of $10^{8} M_{\odot}$ and an accretion rate of $\dot m = 0.35$ in 
Eddington units.
The heating of the gas was always assumed to be proportional to the gas 
pressure,
since the radiation pressure instability
in the AGN disks results in much stronger changes in disk density and thickness
between the hot and cold states. Therefore, the evolution of an
unstable disk is much more 
complex from the computational point of view, i.e. it requires 
a substantial fine-tuning of the time step, and at the moment it is beyond
the scope of our computer facilities. 

\begin{figure}
   \centering
  \includegraphics[width=8.5cm]{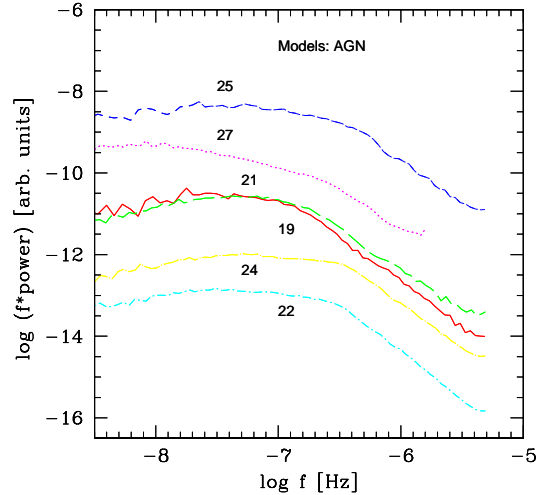}   
      \caption{Power spectra calculated from numerical models
for supermassive black hole in AGN, 
with the prescriptions $A$, $B$, and $C$
for mass evaporation rate (cf. Eq. \ref{eq:mz}
and Tab. \ref{tab:numeric}); the labels on separate curves refer to the
number of the model as given in the Table.
              }
         \label{fig:pds_AGN}
   \end{figure}

The Figure \ref{fig:pds_AGN} shows the
resulting power spectrum density for 6 models with various prescriptions
for the mass evaporation rate and magnetic cell size.

The influence of the parameterization of magnetic cell's size ($a, b$ or $c$)
on the PSD slope 
is less crucial than the role of mass evaporation parameterization 
($A, B$ or $C$).  In general, the cell's size $\Delta r $ 
proportional to $\sqrt{r}$ (case $c$)
gives either similar or slightly flatter slopes than $\Delta r \propto r$ 
(case $b$), while the flattest slopes are produced in the models with 
cell's sizes $\Delta r \propto H$ (case $a$).
 The slope of the model 27 is exceptionally steep for a class A model but this is connected with the frequency location of the slope determination; a small shift towards lower frequencies brings this model back to the general systematic trends.

\begin{figure}
   \centering
   \includegraphics[width=8.5cm]{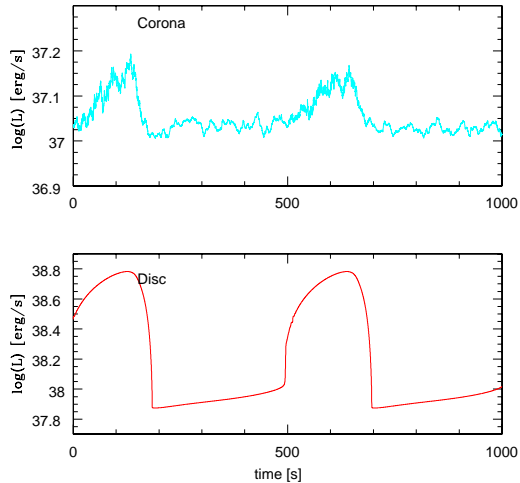}
      \caption{Exemplary fragment of the disk and corona lightcurves obtained
      from the evolutionary model of an unstable disk 
(heating proportional to $P_{\rm tot}$; model 04 in Table \ref{tab:numeric}). 
The model parameters: $M=10 M_{\odot}$, $\dot m_{\rm out} = 0.26$,
$\beta_{\rm S} = 0.5$.
      The total duration of the simulation is 46 ks.
              }
         \label{fig:curves04}
   \end{figure}

\begin{figure}
   \centering
   \includegraphics[width=8.5cm]{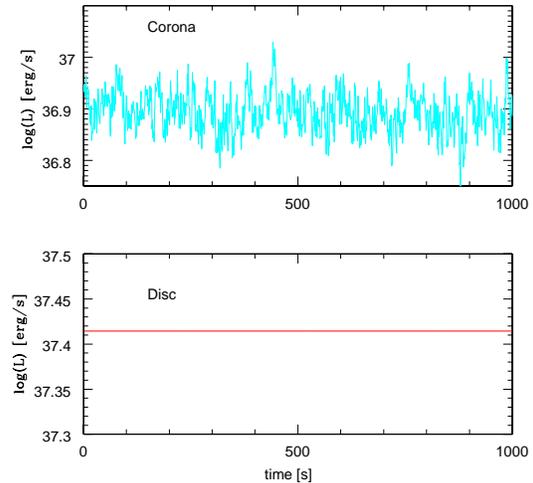}
      \caption{Exemplary fragment of the disk and corona lightcurves obtained
      from the evolutionary model of a stable disk 
(heating proportional to $P_{\rm gas}$; model 13 in Table \ref{tab:numeric}). 
The model parameters: $M=10 M_{\odot}$, $\dot m_{\rm out} = 0.026$,
$\beta_{\rm S} = 1.0$.
      The total duration of the simulation is 48 ks.
              }
         \label{fig:curves13}
   \end{figure}

\begin{figure}
   \centering
   \includegraphics[width=8.5cm]{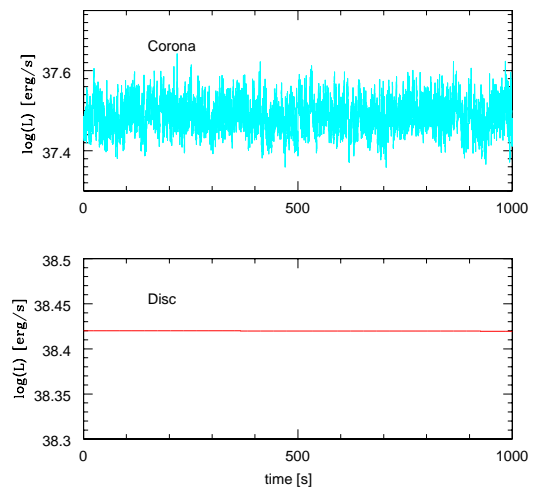}
      \caption{The same as in Fig. ~\ref{fig:curves13}
but for the model 18 in Table \ref{tab:numeric} ($P_{\rm gas}$ heating; 
accretion rate $\dot m = 0.26$, $\beta_{\rm S} = 0.5$ ).
      The total duration of the simulation is 45 ks.
              }
         \label{fig:curves18}
   \end{figure}

The exemplary lightcurves of the disk and the corona 
of Galactic black hole binaries are shown in 
Figures ~\ref{fig:curves04}, ~\ref{fig:curves13} and
~\ref{fig:curves18}. 
In Fig.  ~\ref{fig:curves04} we show the results for the 
unstable disk evolution 
(04 in Table \ref{tab:numeric}), with a large mean accretion
rate and disk heating proportional to the total pressure.
This model results in strong disk luminosity outbursts.
The coronal luminosity follows these long outbursts, 
and the short variability fluctuations due to the magnetic dynamo is 
superimposed on the long term evolution.
The magnetic dynamo parameter, $\beta_{\rm S}$, was assumed to be 0.5,
and nevertheless the coronal fluctuations had a substantial amplitude.

In Figs.  ~\ref{fig:curves13} and ~\ref{fig:curves18}
we show the lightcurves resulting from a stable disk model (13 and 18
 in Table \ref{tab:numeric}), with a small mean accretion
rate and disk heating proportional to the gas pressure.
The models differ with respect to the
prescription for the mass evaporation rate and cell size 
(A-c and C-c, respectively).
The disk emission is practically constant but the
coronal emission shows  modulation due to the dynamo process.
Here the magnetic parameter, $\beta_{\rm S}$, had to be taken
large, in order to produce significant coronal variability.

\begin{figure}
   \centering
   \includegraphics[width=8.5cm]{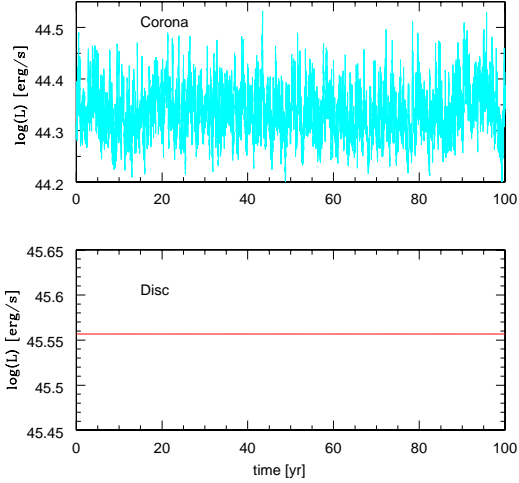}
      \caption{Exemplary fragment of the disk and corona lightcurves obtained
      from the evolutionary model of a stable AGN disk 
(heating proportional to $P_{\rm gas}$; model 24 in Table \ref{tab:numeric}). 
The model parameters: $M=10^{8} M_{\odot}$, $\dot m_{\rm out} = 0.35$,
$\beta_{\rm S} = 0.1$.
      The total duration of the simulation is 415 yrs.
              }
         \label{fig:curves24}
   \end{figure}

In Figure ~\ref{fig:curves24} we show the lightcurves for the 
AGN evolution (model 24 in the Table).
All the models calculated for AGN disks assumed 
the gas pressure domination. The key element of the models
was the prescription for the mass evaporation rate
and the cell's size: the model ``C-a'' gave a 
significant variability in short timescales.

\section{Discussion}
\label{sect:discussion}

Power spectrum density (PSD) of the X-ray emission of Galactic black holes and
active galactic nuclei is usually well approximated by a single 
power law with a slope
$\sim 1$, turning down at high frequencies, if the source is in the Soft State.
This means that a broad range of frequencies contribute to the lightcurve, 
although most of the energy is dissipated close to the black hole, where the 
highest frequencies should dominate. In the present paper we propose 
a two-layer accretion flow - an accretion disk with an accreting corona - as a 
solution to the problem.

The accreting corona model was considered before in a number of papers 
(\. Zycki et al. \cite{zycki95}, Witt et al. \cite{witt97}, 
Janiuk et al. \cite{janiukadv}, R\' o\. za\' nska \& 
Czerny \cite{rozanska2000},
Meyer-Hofmesister \& Meyer \cite{meyer2001}, Misra \& Taam \cite{misra}, 
Meyer-Hofmeister et al. \cite{ema2005}). The model has several known 
advantages.
For example, it can explain the phenomenon known as the hysteresis effect, 
i.e. for a given source luminosity the spectra are softer when the count rate 
is decreasing, and harder when the count rate is increasing (Smith et al. \cite{swank}, Meyer-Hofmeister et al. \cite{ema2005}, Pottschmidt et al. 
\cite{katja2006}), as well as the time delays of the hard X-ray band with 
respect to the soft X-ray band in GRS~1915+105 (Janiuk \& Czerny 
\cite{janiuk2005}).

In this paper we considered the propagation of perturbations in the 
accretion flow 
created locally by the magneto-rotational instability. We followed 
the general approach 
of King et al. \cite{king} and Mayer \& Pringle \cite{mayer}, but we 
incorporated their 
scheme into our two-layer flow. Since the viscous timescale in the 
corona is short, 
local perturbations affecting the coronal accretion rate propagate 
without significant smearing. 
Therefore, the two-flow model offers also in a natural way a physical 
realization of 
the suggestion made by Lyubarskii (\cite{Lyubarskii}) that the broad PSD 
is produced by
the inward propagation of the perturbations created at various disk radii,
so the inner dissipation region preserves the memory of all fluctuations.
Propagation through the disk, considered originally by Lyubarskii (1997), 
and
later explored by King et al. (\cite{king}) and Mayer \& Pringle 
(\cite{mayer}), 
requires rather
long local magnetic timescale; otherwise the fluctuations are washed out in 
course of propagation. In the case of disk/corona coupling the magnetic field
fluctuations result in modulations of the coronal flow, and the viscous 
timescale in the corona is much shorter than in the disk, so the conditions
to preserve the fluctuations are easier to satisfy.

The PSD spectra based on this model were calculated numerically, 
but we also provided
a semi-analytical insight.
Below, we discuss the theoretical predictions of our model, and then we
compare them with the observational data for the Soft State X-ray binaries 
and Narrow Line Seyfert 1 galaxies.

\subsection{Model predictions} 

In the modeling, we use a couple of different scaling laws for the magnetic 
dynamo cell: the cell size is either assumed proportional to the
radius of the disc, or its square root, or to the disc thickness. This last 
option is widely believed to be the most physical one since the 
magneto-rotational instability operates in a scale of the disk thickness, or
a fixed fraction of it, given effectively by the parameter $\alpha$
(e.g. Tout \& Pringle \cite{tout92}, Hawley et al. \cite{hawley95}, 
Balbus \& Hawley \cite{balb98}). On the other hand, larger scale fields can 
be created either
by inverse cascade mechanism and reconnection (Livio \cite{livio1996}, 
Tout \& Pringle \cite{tout},
King et al. 2004), or by field generation in the corona.

\begin{figure}
   \centering
   \includegraphics[width=8.5cm]{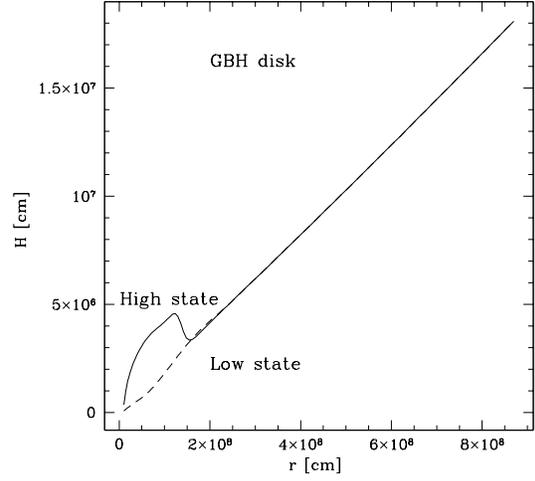}
      \caption{The thickness of the disk as a function of radius
in the unstable model of GBH ($P_{\rm tot}$ heating). The solid and dashed lines show the
H(r) profile in the hot and cold states, respectively.
The black hole mass is $10 M_{\odot}$ and accretion rate is $\dot m=0.26$.
             }
         \label{fig:hdisk_gbh}
   \end{figure}

\begin{figure}
   \centering
   \includegraphics[width=8.5cm]{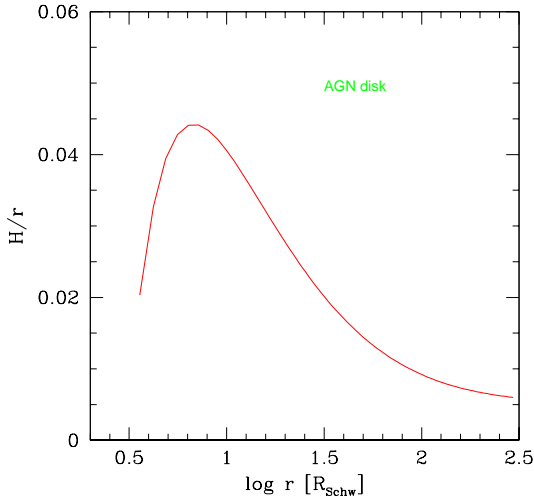}
      \caption{The ratio of the disk thickness to the radius as a function of radius
in AGN ($P_{\rm gas}$ heating). The black hole mass is $10^{8} M_{\odot}$ and the accretion rate
is $\dot m = 0.35$.
             }
         \label{fig:hdisk_agn}
   \end{figure}

When we assume that the characteristic dynamo cell is determined by the disk 
height, 
the resulting power spectrum strongly depends on its radial profile, 
as already noticed by Nowak \& Wagoner (\cite{nowak95}).
In particular, the result of this  assumption 
is such that the models with stable and unstable disks in GBHs, 
as well as the model for the AGN disk, differ substantially from each other
even for the same prescriptions for the mass evaporation rate. 
%if the magnetic cell size scales with the disk thickness.
This is because the disk thickness increases locally in the unstable zone of
the radiation pressure dominated disk in GBH (cf. Fig \ref{fig:hdisk_gbh}). 
In AGN gas pressure dominated models the disk thickness in the outer rises slowly in the outer parts (cf. Fig \ref{fig:hdisk_agn}).
 The slopes of the models (16) and (22) for a GBH and an AGN as given in Table~\ref{tab:numeric} are both similar but this is the coincidence. If we calculate the AGN slope in the frequency range of $\log f $ between -8.8 and -7.8, we obtain much flatter slope of 0.43. When we assume geometrically fixed dynamo cells (b or c) the differences between the models remain. In particular, the conclusion based on analytical models that for Cb $P_{gas}$ models the slope is 1.0 both for GBH and AGN is not supported by the numerical computations: the slope is 0.76 (model 26) and 0.93 (model 25), but both spectra show certain curvature.

The prescription ``B'' for the mass evaporation rate gives always
much too strong emission in the corona (e.g. for 10 $M_{\odot}$ black hole $\log L_{\rm disk} \sim 38.5$, and
$\log L_{\rm cor} \sim 39$),
which in these models exceeds the disk luminosity by a factor of $\sim 3$.
This is not consistent with the Soft State spectra being dominated by the disk
component. Strong outflow from the disk to the corona in such models leads
to the corona being optically thick for scattering, while the disk surface 
density drops considerably in comparison with prescription ``A'' models for 
the same accretion rate, although it is never completely evaporated. We conclude that the assumption of the light speed 
instead of the 
Alfven velocity in energy flux transported to the corona does not lead to
acceptable models.

The evaporations laws ``A'' and ``C'' give disk-dominated models, and ``C'' 
(corresponding to modified inner boundary condition) gives always much flatter 
power spectra. Within the frame of these models, the magnetic dynamo parameter 
$\beta_{\rm S}$ plays also an important role.
In the case of a stable disk in GBH and in AGN, it usually (models with $P_{\rm gas}$ heating)
has to be taken equal to 1.0, so that the dynamo can produce the
substantial amplitude of X-ray (coronal) variability, supported by the 
observations of the Soft State sources
(e.g. Treves et al. \cite{treves88}, Nowak et al. \cite{nowak01}, Churazov et al. \cite{chura2001}, Axelsson et al. \cite{axelsson} 
for GBH; Markowitz \cite{3783power}, Uttley \& McHardy \cite{ut5506}, Shemmer et al. \cite{shemmer} 
for AGN).

The $\beta_{\rm S}=1$ means that the global (large scale) magnetic field 
forms effectively and supports the energy transport between the disk and 
corona. 
In the unstable GBH disks, the stochastic fluctuations of coronal emission
are large even for the values of $\beta_{\rm S}$ smaller than 1.0,
i.e. we can assume it to be 0.5; otherwise, the corona is too strong.
Note, that we are using here a constant value of the $\beta_{\rm S}$ parameter,
but other possibilities have also been recently explored. For instance, 
Czerny et al. (\cite{universal}) and Tout \& Pringle (\cite{tout}) suggested 
that $\beta_{S}$ is proportional to $H/r$. Such a prescription steepens or
flattens the PSD, depending on the radial profile of $H/r$.

Finally, we stress that the mass of the central
black hole also plays the role, and the PSD spectra corresponding to the accreting
supermassive black holes are usually steeper in full numerical simulations of
disk-corona evolution, in comparison to the analytical estimations. 
This is because the disk properties, when calculated
properly, do depend on the black hole mass in a complex way. 

Analytically, the evaporation rate is proportional
to $ B^{2}_{\rm z} v_{\rm Alvf}$ 
(see Sect.~\ref{sect:res1}), 
which, in the stable accretion disk,  depends on the 
radius as $r^{-3}$, regardless of the central mass.
However, in the numerical simulation, when the coupling with the corona is 
included, this relation is modified. We find that for GBH disk the 
(averaged, i.e.
 not including the random changes) slope
of $\dot m_{\rm z}(r)$ is $\sim -3.28$, while for AGN it is
$\sim -2.89$ in the main inner part of the disk 
(up to $\log r \sim 15.8$ cm for $M_{\rm BH}=10^{8} M_{\odot}$;
cf. Fig. \ref{fig:hdisk_agn}). Moreover, in the outer disk part 
(which is much more geometrically thick), the slope of this function changes
dramatically and becomes positive: $\sim 4.3$.
These properties are the reason for the systematically steeper PSD slopes
in AGN, since in those disks the longer timescales are more 
pronounced than in Galactic sources.

King et al. (2004) did not find any dependence on black hole mass, as 
these authors assumed that all the quantities scale with the BH mass in a 
self-similar way and the ratio $H/r$ is constant.
Mayer \& Pringle (2006) did not study a difference between the PSD spectra 
in stellar mass and supermassive black holes, and the dynamo 
and power spectra were calculated only for a fixed mass of 10 $M_{\odot}$. 
These authors discuss, however, the systematic differences 
in the instability picture. They find that
for higher mass the disk becomes more unstable due to the larger range of
accretion rates which are in the unstable regime.
%2. they calculate disk limit cycle for $10^{6} M_{\odot}$, and with v. small number 
%of radial points in the grid. The Z-bump instability is an additional source of flickering. 
%For higher mass, the unstable strip is too large in radius and they do not calculate this.

\subsection{Observed PSD of the sources in the Soft States}

A few Galactic sources are permanently in the Soft State, many more
sources (several transient X-ray sources, in particular) exhibit temporary 
Soft States (for a review, see Remillard \& McClintock \cite{remillard06}).

\begin{table*}
\caption{Power spectra of Galactic Black Holes in Soft State}             
\label{table:gbh}      
\centering          
\begin{tabular}{c c c c  l }     % 5 columns 
\hline\hline       
 object & low frequency slope & high frequency slope & frequency break & reference\\ 
\hline                    
Cyg X-1 & 0.92 & -  & $\sim 10 $ Hz & Axelsson et al. \cite{axelsson} \\
XTE J1650-500 & $\sim 0.06$ & $\sim  1.4 $ &  $\sim 2$ Hz & Done \& Gielinski \cite{dg2005} \\
GRS 1915+105  & $1.3 \pm 0.1 $ & - & $ > 10 $ Hz & Rao et al. \cite{rao} \\
GRS 1915+105  & $1.3 \pm 0.1 $ & $1.61 \pm 0.02$ & $ 1 $ Hz & Rao et al. \cite{rao} \\
GRS 1915+105  & $1.25 \pm 0.02 $ & - & $ > 10 $  Hz & Rao et al. \cite{rao} \\
LMC X-1  & $ \sim 1 $ & - & $ 0.2  $  Hz & Nowak et al. \cite{nowak01} \\
\hline                  
\end{tabular}

\end{table*}

The shape of the PSD was studied for a few sources in the Soft State (see 
Table~\ref{table:gbh}). In most of the sources the power spectrum is broad, 
with a slope 0.9 - 1.3. The power spectrum of XTE J1650-500 is an exception
since in this case the dominating contribution to the power comes from a 
relatively narrow range of frequencies, around and below $\sim 2$ Hz.

\begin{figure}
   \centering
   \includegraphics[width=8.5cm]{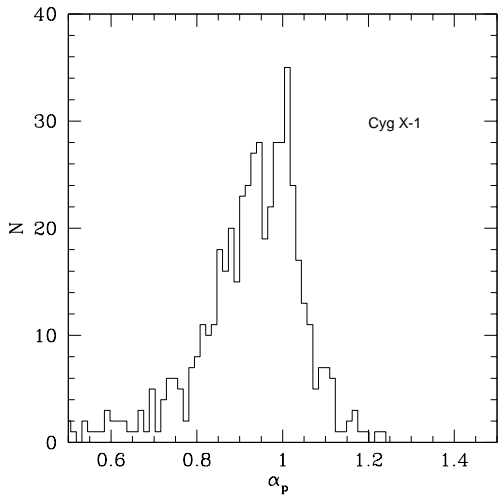}
      \caption{The distribution of the slopes of the power law 
      component in the power
      spectrum of Cyg X-1 in the Soft State (data taken from table~1 
      of Axelsson et al. \cite{axelsson})
             }
         \label{histogram}
   \end{figure}

The most detailed analysis of the Soft State was done for 
Cyg X-1 by Axelsson et al. (\cite{axelsson}). They 
analyzed 614 power spectra by decomposition into two Lorentzians 
and a 
power law component with high frequency cut-off 
$P(f) = af^{(-\alpha_p)}\exp(-f/f_c)$.
The distribution of the power law indices $\alpha$ for 523 successful fits 
is shown in 
Fig.~\ref{histogram}. The mean slope is $p=0.92$, with the dispersion of 0.18, 
and the mean high energy cut-off is 10.0 Hz, with the dispersion 9.4 Hz.

%----------------------------------------------------------- 
%
%______________________________________________________________

\begin{table*}
\caption{Power spectra of Narrow Line Seyfert 1 galaxies}             
\label{table:agn}      
\centering          
\begin{tabular}{c c c c  l }     % 5 columns 
\hline\hline       
 object & low frequency slope & high frequency slope & frequency break & reference\\ 
\hline                    
MCG -6-30-15 & $0.8^{+0.16}_{-0.4}$ & $1.98^{+0.40}_{-0.32}$ & $6.0^{+10}_{-5} \times 10^{-5}$ Hz & McHardy et al. \cite{mcgpower} \\
NGC 4051 & 1.1 (fixed) & $2.03^{+0.7}_{-0.4}$ & $8.3^{+19}_{-8} \times 10^{-4}$ Hz & McHardy et al. \cite{mchardy2004} \\
Mrk 766  & $\approx 1 $ & $2.8^{+0.2}_{-0.4}$ & $\approx 5 \times 10^{-4}$ Hz &  Vaughan \& Fabian \cite{766power} \\
Ark 564 & $ 1.24^{+0.03}_{-0.04}$ & $1.90^{+0.17}_{-0.14}$ & $2.3^{+0.6}_{-0.6} \times 10^{-3}$ Hz & Papadakis et al. \cite{564power} \\
NGC 5506 & $\sim 1.0 $ & $ > 1.7 $ & $1.3^{+8.3}_{-0.7} \times 10^{-5}$ Hz & Uttley \& McHardy \cite{ut5506} \\
\hline      

\end{tabular}

\end{table*}

Among AGN, it is widely believed that Narrow Line Seyfert 1 galaxies
are high mass analogs of the GBHs in the Soft State (e.g. Pounds et al. 
\cite{pounds95}, McHardy et al. \cite{mchardy_rev}, Uttley \& McHardy \cite{ut5506}, Czerny \cite{czerny_rev}). 
In Table~\ref{table:agn} 
we list objects with determined PSD slopes. In this case the power spectrum is not determined 
so accurately as in the case of Galactic sources, but the overall trend seems to be the same. 
In particular, the slope is about $0.8 - 1.3$ both in the Soft State GBH and in 
NLS1s, although  better measurements is clearly needed. 

Several models considered in our paper satisfy this requirement. Among 
the semi-analytical models, 
examples are: Ab-rad and Ab-gas, as well as Aa-gas and Ca-gas, 
which are roughly consistent with this requirement.
Also, models Cb-rad and Cb-gas certainly provide good representation of the data. 
As for the full numerical models, several small mass models are acceptable (01, 03, 06, 07, 09, 13, 16 and 28 in Table \ref{tab:numeric}),
 and four of the high mass models is satisfactory (19, 21, 22, 25).

\section{Conclusions}
\label{sect:conclusions}

In this article we present the calculations of the time-dependent
accretion disk plus corona model, incorporating the  magnetic dynamo
as a mechanism for the origin of the short-term variability in hard X-rays.
The current model is suitable for the Soft State X-ray binaries, as well as 
their supermassive black hole analogues: Narrow Line Seyfert 1 galaxies.
The radiation spectrum is dominated by the thermal soft component, i.e. 
the accretion disk, which may either be constant in time, or
variable on relatively long timescale due to the instability induced
by the radiation pressure domination. The hard spectral component 
carries less energy, and is produced by the accreting corona above the disk.
The coronal emission is variable on short timescale due to the the rapidly 
changing magnetic field in the accretion disk. The key element of the model
is therefore the coupling between the disk and corona, which due to the 
mass evaporation driven by the magnetic flux allows the 
perturbations generated locally in the disk to be transported to the corona.
Since the viscous timescale in the corona is much shorter than in the disk,
these perturbations are not smeared and a broad power spectrum is
produced. 

By studying our model both analytically and numerically, as well as by the 
comparison with the available observational data, we drive the following 
conclusions:
   \begin{enumerate}
      \item The radiation pressure instability
in the underlying accretion disk leads to steeper power density spectrum of 
its fluctuations, as well as to the larger amplitudes.
      \item The assumption of the light speed instead of the 
Alfven velocity in energy flux transported to the corona does not lead to
acceptable models.
     \item  The PSD spectra never look like a broken power law, they turn off slowly at higher frequencies and frequently show some curvature at low frequencies as well.
      \item The PSD spectra corresponding to the accreting
supermassive black holes for most models are steeper
than for the galactic black holes. This result has been obtained 
in full numerical simulations of
disk-corona evolution, while the analytical estimations 
give PSD independent on BH mass.
   \end{enumerate}

The latter point might seem inconsistent with the AGN data obtained so far, since their
PSD spectra seem to have slopes similar to these of Galactic sources.
However, one should take into account the still poor statistics and large 
errors of these observations. We conclude that further verification of the 
models requires more detailed observations and analysis of  the long-term 
variability in AGN.

\begin{acknowledgements}
We thank to Piotr \. Zycki for helpful discussions. 
Part of this work was supported by grant 
1P03D00829 of the Polish
State Committee for Scientific Research.
\end{acknowledgements}

\section*{Appendix}

In the computations of the time evolution of the disk corona system, 
the energy equation ({Eq.~23) is simplified due to the vertical integration
over the disk height (see e.g. Taam \& Lin 1984). This approach neglects the components
connected with the time evolution of the disk thickness, which means an assumption that
the density decreases significantly at the disk surface.
Here we derive a full energy equation that includes other 
the necessary terms,  e.g. assumes that the vertical distribution of density in the
disk is uniform. Then
we check if our simplified treatment is justified.

The energy equation in its basic for is:
\begin{equation}
T{dS \over dt} \Sigma = Q^+ - Q_-
\end{equation}
where $S$ is the entropy and other terms have the same meanings as above.

From the thermodynamical relation we have:
\begin{equation}
TdS = {P \over \rho}[(12 - 10.5 \beta)dlnT - (4 - 3 \beta)dln \rho]
\end{equation}
and therefore
\begin{equation}
{P \Sigma \over \rho}[(12 - 10.5 \beta){dlnT\over dt} - (4 - 3 \beta) {dln \rho \over dt}] = Q^+ - Q_-.
\label{eq:app3}
\end{equation}
Dividing the above equation by $ P \Sigma/\rho/(12 - 10.5 \beta)$ we have:
\begin{equation}
{dlnT \over dt} = {4 - 3 \beta \over 12 - 10.5 \beta} {dln \rho \over dt} +
{(Q^+ - Q_-) \rho \over P \Sigma (12 - 10.5 \beta)},
\end{equation}
and substituting $\rho$ with $\Sigma/H$, this turns into:
\begin{equation}
{dlnT \over dt} = {4 - 3 \beta \over 12 - 10.5 \beta} {dln \rho \over dt} +
{(Q^+ - Q_-) \over P H (12 - 10.5 \beta)}.
\end{equation}

Since we use $d lnT = dT/T$, the above equation is divided by $T$.
However, using $d\ln\rho = d\ln\Sigma$ is an simplification that we have done in the numerical calculations
presented above.
To be strict, we should take rather:
\begin{equation}
\rho = {\Sigma \over H}
\end{equation}
which is equivalent to:
\begin{equation}
\ln \rho = \ln \Sigma - \ln H
\end{equation}
and 
\begin{equation}
d\ln\rho = d\ln\Sigma - d\ln H
\label{eq:app10}
\end{equation}
In order to have $d lnH$ as a function of $T$ and $\Sigma$, we take
the hydrostatic balance equation:
\begin{equation}
{P \over \rho} = \Omega^2 H^2
\end{equation}
and
\begin{equation}
P = \Omega^2 H^2\rho = {\Omega^2 H^2 \Sigma \over H} = \Omega^2 H \Sigma
\end{equation}

The pressure is a sum of gas and radiation pressures, and therefore:
\begin{equation}
P = {k \over m_p \mu}\rho T + {1 \over 3}aT^4 = {k \over m_p \mu} {\Sigma T \over H} + {1 \over 3}aT^4
\end{equation}
From the two equations above, we have:
\begin{equation}
{k \over m_p \mu} {\Sigma T \over H} + {1 \over 3}aT^4 = \Omega^2 H \Sigma
\end{equation}

Now we differentiate our equation, taking into account that the first term above
is equal to $\beta P$, and the second term is  $(1 - \beta)P$.
Starting with:
\begin{eqnarray}
{k \over m_p \mu}T {d\Sigma \over H} & - {k \over m_p \mu}\Sigma T {dH \over H^2}+ {k \over m_p \mu} \Sigma {dT \over H} + {4 \over 3} aT^3 dT  = \nonumber \\
& d(\Omega^2) H \Sigma + \Omega^2 dH \Sigma + \Omega^2 H d\Sigma
\end{eqnarray}
and after a couple of arithmetical operations we have:
\begin{equation}
(\beta - 1) d\ln \Sigma + (4 - 3 \beta) d\ln T - d\ln (\Omega^2) = (1 + \beta) d\ln H 
\end{equation}
From the above equation we obtain $d lnH$ and put it into the Eq. \ref{eq:app10}, which gives:
\begin{equation}
d\ln\rho = {2 \over 1 + \beta}d \ln \Sigma - {4 - 3 \beta \over 1 + \beta} d \ln T + {1 \over 1 + \beta} d\ln (\Omega^2).
\end{equation}
This relation is now substituted into the equation \ref{eq:app3}:
\begin{eqnarray}
{P H}[(12 - 10.5 \beta) d\ln T & - {4 - 3\beta\over 1 + \beta} [2 d\ln \Sigma - (4 - 3 \beta) d\ln T + \nonumber \\
& dln (\Omega^2)]] =  Q^+ - Q_-
\end{eqnarray}
and after a few transformations we have:
\begin{equation}
dlnT = {2(4 - 3 \beta)\over A_B} dln \Sigma + {(4 - 3 \beta)\over A_B}dln\Omega^2) + {(1 + \beta)(Q^+ - Q_-) \over P H  A_B}
\end{equation}
where:
\begin{equation}
A_B = [(12 - 10.5 \beta)(1 + \beta) + (4 - 3\beta)^2]
\end{equation}

Finally, we delogarythimize the equation:
\begin{equation}
dT = {2T(4 - 3 \beta)\over A_B \Sigma} d \Sigma + {T(4 - 3 \beta)\over A_B}dln\Omega^2) + {T(1 + \beta)(Q^+ - Q_-) \over P H A_B}
\end{equation}
and substitute the full time derivatives of $dT$ and $d \Sigma$ with partial ones:
\begin{equation}
dT ={\partial T \over \partial t} + v_r {\partial T \over \partial r} 
\end{equation}
\begin{equation}
d\Sigma ={\partial \Sigma \over \partial t} + v_r {\partial \Sigma \over \partial r} 
\end{equation}

This equation is more complicated than the one used in our numerical calculations. Therefore, let us now check, how the numerical results depend
on the adopted formula for $d T/dt$.
We performed several  test runs with the above relation  and we found that 
the results only slightly differ from the
ones with the simplified formula used throughout this work. 
The test runs were performed for models 01, 04, 07 and 19 as in the
Table \ref{tab:numeric}.
In general, for the models in which the disk was stable (07 and 19), 
the value of its constant luminosity was exactly the same in the test runs 
as in the Table.
For the models in which the disk was unstable due to the radiation 
pressure (01 and 04), the amplitudes of the luminosity outbursts and 
their durations changed by about 0.1\%. The same range of 
change occurred for the amplitudes of the coronal variability in all of the
tested models.
For the slopes of the power density spectra, they changed by 9\%,
3\%, 2\% and 8\%, for the test models 01, 04, 07 and 19, respectively.
These differences were most probably due to the numerical reasons
(the time step used in our model is variable, depending on the rate 
of change of temperature and density in the disk).
We conclude, that the approximation used in case of the energy 
equation in the time dependent numerical calculations is viable and does 
not influence our results
in a great deal.

\end{document}